\documentclass[sigconf,screen,nonacm]{acmart}

\AtBeginDocument{%
  }

\usepackage{xcolor}
\usepackage{todonotes}
\usepackage{amsmath}
\usepackage{graphicx}
\usepackage{subcaption}
\usepackage{printlen}
\usepackage{wrapfig}
\usepackage{booktabs}      
\usepackage{makecell}      
\usepackage{array}         
\usepackage{caption}
\usepackage{multirow}

\usepackage{listings}
\lstdefinestyle{prompt}{
    basicstyle=\ttfamily\tiny, 
    breaklines=true,
    columns=fullflexible,
    keepspaces=true,
    frame=single,
    framesep=5pt,
    backgroundcolor=\color{gray!8},
    literate={–}{{--}}1 {—}{{---}}1 {→}{{$\rightarrow$}}1 {×}{{$\times$}}1
}

\begin{document}

\title{Whose Safety? Addressing Anti-Blackness in the Age of AI Safety}

\author{Mírian Silva}
\authornote{This work was supported and primarily conducted during a research fellowship at the Black in AI Safety \& Ethics (BASE). Corresponding author: mirianfrsilva@gmail.com}
\affiliation{%
  \institution{Black in AI Safety \& Ethics}
  \city{Belo Horizonte}
  \country{Brazil}
} 
\email{}

\author{Serena Oduro}
\affiliation{%
  \institution{Data \& Society Research Institute}
  \city{New York}
  \country{USA}
}
\email{}


\begin{abstract}

“AI safety” has grown from a niche contingent of actors into the dominant framing for artificial intelligence (AI) assessment and evaluation, overshadowing AI ethics. Through political and monetary influence, AI safety has shifted political focus within the US and abroad. The focus has shifted away from harms that currently affect the public toward threats framed as “existential risks”. We argue that safety cannot be achieved while anti-Blackness is left unaddressed. We ground this argument in Sylvia Wynter’s critique of the “overrepresentation” of Man: the modern, Western conception of the Human that is predicated on the subjugation and othering of Blackness. To analyze the state of addressing anti-Blackness in the age of AI safety, we conduct a two-stage, language model-assisted content analysis of \textbf{69,290 accepted main-track} papers from ten leading machine learning (ML), natural language processing (NLP), and AI ethics conferences (2020–2025). Safety research grew from 0.5\% of accepted papers in 2020 to 6.4\% in 2025, overtaking ethics. Yet 94\% of safety papers rely on automated metrics, and only 11\% of safety papers mention harms that affect society directly, compared with 78\% of ethics papers. Of the 2,182 safety papers, 8 mention race; none make it central, and none name Black or African-diaspora communities. Across all 254 identified as race-focused papers, only 3 describe racial harm in safety terms, and 55\% appear at three ethics venues that publish 2.5\% of the corpus. We offer this interdisciplinary examination of AI safety to argue that a just technological future depends on confronting anti-Blackness directly.


\end{abstract}



\maketitle

\section{Introduction}
\label{sec:intro}

In recent years, “AI safety” has moved from a niche technical concern to an epistemic culture that has grown to dominate the field of artificial intelligence (AI) governance and evaluation \cite{ahmed2024field}. It has shifted political, organizational, and research objectives away from regulating and evaluating harms toward the public to preventing catastrophic risks from deploying automated AI systems \cite{nist2024network, govca2025sb53, nygov2025raise}. Frontier developers, government safety institutes and research nonprofits now decide which risks are benchmarked, which metrics count as evidence of progress, and which failures require action \cite{bengio2026internationalaisafetyreport,ahmed2024field, de2026WhatIsSafety, weidinger2023sociotechnical}. In doing so, they define what the field recognizes as safety.


That definition centers safety on preventing dangerous capabilities, specifically chemical, biological, radiological, and nuclear (CBRN) uplift, cyber-offense, and loss of control \cite{bengio2026internationalaisafetyreport, phuong2024evaluatingfrontiermodelsdangerous}. Under this paradigm, catastrophic misuse and alignment failures count as safety risks, while anti-Blackness, racism, and other forms of discrimination, along with ecological crises, fall under ethics and fairness agendas and are treated as less significant than dangerous capabilities or the prospect of AI achieving superintelligence.

Yet, anti-Blackness' catastrophic impacts on the Black community \textit{and} our modern conception of being are present whether or not AI safety deems it to be. Sylvia Wynter critiques Man's overrepresentation in the modern Western conception of humanity; Man is the white, bourgeois man, and all others, including Black and marginalized people, are positioned as the Other \cite{Wynter2003}. Anti-Blackness provided the historical justification for slavery and colonialism, and it continues to support a Western society that justifies, relies upon, and accepts Black death. \citet{GebruTorres2024_TESCREAL} trace much of the AI safety movement’s intellectual lineage to first-wave eugenics. Anti-Blackness being deemed outside the scope of safety is not an accident, then, but a feature of the communities’ intellectual roots and goals that are deeply connected to anti-Blackness. 


We argue that the scope of AI safety works as what \citeauthor{Wynter2003} calls a \textit{descriptive statement} of the Human: it decides whose well-being stands for all of humanity, and so, in the context of AI, whose harm counts as catastrophe and whose counts as bias (\S~\ref{sec:background}). Rather than assume this, we test it.

We assess the state of addressing anti-Blackness in AI today through analyzing 69,290 (69,344 collected) papers accepted between 2020–2025 at ten leading machine learning (ML), natural language processing (NLP), and AI-ethics venues using a two-stage, language model-assisted content analysis that codes each paper’s domain, subdomain, methods, and evidence for their treatment of racial harms (\S~\ref{sec:method}). We organized it into three parts, pairing each part with an expectation drawn from \citet{Wynter2003}’ critique:


\begin{enumerate}
    \item \textbf{Trajectory} (where the field is going). As safety grows, does research on anti-Black harm grow with it, stay in place, or concentrate in venues that safety research does not enter? We expect it to stay out of the venues where safety is published; the expectation fails if race-focused research rises at ML and NLP venues alongside safety.
    \item \textbf{Practice} (methods and evidence). Do the methods and kinds of evidence that safety research relies on match the methods and evidence through which anti-Black harm has been documented? We expect safety to rely on evidence measured on models; the expectation fails if safety uses participant, deployment, or audit evidence at rates near those of ethics research.
    \item \textbf{Concepts}. When published research addresses harm to people, and racial harm to Black and African-diaspora communities in particular, is that harm placed inside or outside the safety frame? We expect racial harm to be framed as bias, not safety, even when safety's own interventions cause it; the expectation fails if such research describes the harm in safety terms.
\end{enumerate}

Due to the growth of the AI safety movement, we use research papers as a site to analyze the current state of addressing anti-Blackness, because current trends may likely become the default perspectives and methods for AI evaluation as a whole. In a political and technological movement that actively contributes to anti-Blackness, especially by deeming addressing anti-Blackness to be out of scope of evaluation metrics, we are interested in the actual presence or claim within papers within the past six years to address anti-Blackness. We want to highlight that even if anti-Blackness is named, that does not mean it is being addressed adequately, nor that solutions will actually improve Black well-being~\cite{Birhane2022_MarginsAIEthics}. 

Through our content analysis, we found that all three expectations hold (\S~\ref{sec:findings}): AI safety research grew from 0.5\% to 6.4\% of accepted papers, overtaking ethics, yet none of its 2,182 papers name Black or African-diaspora communities. \citet{Roytburg2026} show through co-authorship networks \textit{who} is separated across the safety–ethics domains; we examine \textit{what} these domains contain: which harms each side studies, which evidence it accepts, and whose harms fall outside safety. Our contributions are mainly: (1) \textit{A theory-driven test from Black studies.} We read the governing definition of AI safety through Wynter's critique of the overrepresentation of Man, derive three expectations, and state in advance which results would contradict them; and (2) \textit{A measurement of the scope of AI safety}. We measure which harms, methods, and kinds of evidence each field accepts across six years of published papers; 

Section \S~\ref{sec:background} develops the theory and describes how AI safety became dominant; \S~\ref{sec:method}–\ref{sec:findings} present the method and test each expectation; \S~\ref{sec:discussion} discusses the results and what must change for anti-Blackness to be assessed and addressed; and \S~\ref{sec:conclusion}  concludes the paper.

\section{Understanding Anti-Blackness to Understand the Human}
\label{sec:background}

Understanding how anti-Blackness informs modern conceptions of Being is important to understand how, even when there are calls for AI safety, calls to prevent catastrophic risks, and calls to prevent human extinction, Black people are still not protected, and, if anything, our subjugation is at the intellectual root of how these calls were justified. \citeauthor{Wynter2003} argues that the struggle of our times is securing our autonomy when our modern conception of the human, Man, a conception created by and meant to protect Western bourgeois interests, is overrepresented and seeks to secure its own well-being: the well-being of the wealthy, white elite \cite{Wynter2003}. The modern conception of Man distances itself from its former Christian-rooted sense of self. Yet instead of the "mortal/immortal, natural/supernatural, human/the ancestors, the gods/God" distinction through which traditional, Christian societies defined what it meant to be human, the Western, modern, secularized approach also relies on a distinctive other: the human/subhuman distinction \cite{Wynter2003}. Used to justify colonialism and slavery, it positioned Black and Indigenous peoples as the “subhuman other”.

However, being positioned as the “subhuman other” did not merely serve to enact violence towards Black and Indigenous peoples. It fundamentally reoriented our conceptions of Being toward valuing the Rational subject above all, and defined the Rational in opposition to, and to discredit, all Black and Indigenous modes of knowledge and relationships to Being \cite{Wynter2003}. The subjugation of Black and Indigenous peoples is therefore directly connected to the technological, political, and ecocidal crises of our times. We cannot accurately address the threat of artificial intelligence without contending with the underlying conception of Man that has justified the technological, political, and corporate practices that have led to our modern crises.

\subsection{AI Safety: Its Roots and Who It Intends to Save}
\label{sec:ai-safety-roots}

Addressing anti-Blackness is particularly challenging with the rise of AI safety due to the community's deep ties to eugenics and its utopian-apocalyptic fantasies that aim to reify the modern conception of Man. \citet{ahmed2024field} describe the AI safety movement as an epistemic community: alignment, interpretability, evaluation, control, and governance researchers bound by shared commitments to long-term and catastrophic risk, shared methods, and overlapping funding networks. It is a dynamic, yet insular, community and ecosystem that has developed mainly outside of the academy; it has its own terminology, sources that frame its knowledge, and goals. \citet{GebruTorres2024_TESCREAL} argue that the intellectual roots of the AI safety movement are connected to a second wave of the eugenics tradition, connecting prominent thinkers and organizations within the AI safety movement to a set of different, yet overlapping ideologies that undergird the AI safety movement, the TESCREAL Bundle — transhumanism, Extropianism, singularitarianism, cosmism, Rationalism, Effective Altruism, and longtermism \cite{GebruTorres2024_TESCREAL}. 
The TESCREAL bundle relies on first-wave eugenicist conceptions of IQ and intelligence, and most of its proponents share the ultimate fantasy of creating artificial superintelligence that is intelligent and rational beyond human ability.


Yet \citet{GebruTorres2024_TESCREAL} argue that, like Christianity and other religions, the TESCREAL bundle relies on a dichotomy of utopia and apocalypse. The utopian fantasy of “paradise-engineering” comes with an apocalyptic threat: a technology powerful enough to fundamentally alter human and cosmic existence would also be strong enough to destroy humanity itself \footnote{https://www.hedweb.com/}. However, the threat of apocalypse has not stopped members of the TESCREAL bundle from developing AGI or artificial superintelligence (even if there have been temporary calls to stop AGI development); instead, it animates their quest to develop it first. 


While the TESCREAL bundle often purports utopian fantasies that liberate all of humanity, just as Wynter theorizes, this conception of Man does not actually include Black and marginalized peoples. The group roots their beliefs in consequentialism and “total utilitarianism”, believing that they should be working toward preventing the extinction of the human race itself and casualties of catastrophic proportions; Issues that impact sub-groups, such as Black people, are not deemed to be existential under their calculus. When Max Tegmark, cofounder of the Future of Life Institute, was asked about the dangers technology presents to the public, such as biometric surveillance, Tegmark said: “Extinction is not something in the very distant future ... And once we’re all extinct, you know, all these other issues cease to even matter” \cite{democracynow2023ai}. When \textit{Rolling Stone} similarly asked Geoffrey Hinton, often deemed to be the “Godfather of AI”, about the risks posed by large language models that Timnit Gebru wrote in a paper \cite{Bender2021Gebru_StochasticParrots} that led to her being fired by Google, he stated: “I believe that the possibility that digital intelligence will become much smarter than humans and will replace us as the apex intelligence is a more serious threat to humanity than bias and discrimination” \cite{oneil2023women_tried_warn}. As \citet{ahmed2024field} note, the problem with this equation is that what is deemed to be catastrophic depends on the person defining what catastrophic \textit{to them} is. It is no surprise that a field deeply connected to eugenics and aiming to create a being more rational than humans does not view anti-Blackness, discrimination, or any societal harm as catastrophic enough to address. Yet, \citeauthor{Wynter2003}'s conception of Man shows that anti-Blackness was created to justify a colonial frame of Being meant to completely reorient human life, knowledge-making, global capital, labor, and land to serve a small group of elites. Through a calculus rooted in \citet{Wynter2003}, anti-Blackness is, indeed, catastrophic. 

\textbf{Frontier frameworks.}
\label{sec:frontier-framweworks}
AI safety's calculus of catastrophic harms has shaped the frameworks that frontier developers publish, usually called Frontier Safety Frameworks (FSF) or Responsible Scaling Policies (RSP). These frameworks emerge from internal design rather than democratic deliberation or community consultation \cite{metr_2025_common_safety_policies, buhl2025emergingpracticesfrontierai}. Despite superficial differences in naming and structure, they converge on CBRN weapons, offensive cyber operations, and loss of control, each with triggers that prevent models from crossing critical capability thresholds \cite{bengio2026internationalaisafetyreport, karnofsky2024_risk_reduction}. Most do not include triggers for discriminatory and racial harm, and they lack a threshold to activate safeguards when a model shows disparate performance across racial groups, reinforces stereotypes at scale, or systematically disadvantages Black users in consequential domains such as housing, employment, or the criminal legal system \cite{bengio2026internationalaisafetyreport}. These harms are architecturally absent from the conditional risk logic that governs when and why companies must act. Two frameworks in the report are exceptions: \textit{NVIDIA} covers “unlawful discrimination at scale”, which escalates to mandatory review by an independent AI ethics committee \cite{nvidia2025_fsf}, and \textit{Cohere} covers “harm in ordinary, non-malicious use”, including discriminatory outcomes \cite{cohere2025_fsf}. These frameworks decide which evaluations get built, and evaluations shape which research is fundable and publishable. The scope codified in policy is therefore also a scope in research practice, which we measure in \S~\ref{sec:method}--\ref{sec:findings}.

\subsection{Addressing Anti-Blackness: Discrimination and Beyond}
\label{sec:interdisciplinarity}

Due to the catastrophic effects of anti-Blackness, we argue that addressing anti-Blackness necessitates, but also extends beyond, assessing for discrimination. The anti-Black violence and harms that AI systems enact are well documented, from intersectional error disparities in face analysis \cite{BuolamwiniGebru2018} and under-referral of Black patients by a deployed health algorithm \cite{Obermeyer19_dissecting_bias} to covert, dialect-based racism in language models \cite{Hofmann2024_ai_racist_decisions}. Anti-Blackness structures not only these outcomes, but also how researchers and engineers approach AI design, development, and deployment\cite{Dancy2022_ai_blackness}. 
\citet{Workman2023} use Wynter to argue that the AI development process, from problem identification to integration and monitoring, frequently reinforces the modern conception of Man and we extend this line of work to the field that now defines safety.


In \S~\ref{sec:method}–\ref{sec:findings}, we use AI safety and AI ethics research papers as sites to investigate how fields with varying disciplinary rigidity attempt to address anti-Blackness. We compare AI safety to the field of AI ethics due to the rich history of AI ethics bringing together fields across the sciences, social sciences, and humanities to address the societal impacts of AI \cite{Bender2021Gebru_StochasticParrots,fort-etal-2011-last}. AI ethics grew out of empirical work on deployed systems that documented discrimination in face analysis, health care, and search \cite{BuolamwiniGebru2018, Obermeyer19_dissecting_bias, Noble2018_AlgorithmsOppression}, and was institutionalized through venues such as the Association for Computing Machinery's Conference on Fairness, Accountability, and Transparency \cite{Laufer2022_facct}. Its frame is sociotechnical: harms arise from the contexts in which and how systems are built and used, not from the model alone \cite{shelby2023harmtaxonomy}. AI safety and AI ethics share concerns such as model opacity and human oversight \cite{Roytburg2026}, which is why we study them together: both claim to protect people from AI, but comparing AI ethics to AI safety shows the differences between whether and how these fields recognize and address harms in research.   
We measure disciplinary rigidity through the methods and kinds of evidence each field accepts (\S~\ref{sec:practice}).

Yet, we do not intend to position AI ethics as a faultless field. \citet{Birhane2022_MarginsAIEthics} also invoke Wynter to critique AI ethics' origins in Western philosophy; Western philosophy often treats ethical concerns as speculative issues with theoretical answers that may have universal applicability, replicating the representation of Man for the Human. While AI ethics has attempted to address the societal harms of AI, it has often fallen into “traps,” rendering work ineffective, often due to misunderstanding how bias, discrimination, and social factors impact technological development and how technology works and impacts the public on the ground \cite{Selbst2019}. Therefore, we intend to compare the state of AI safety research to AI ethics not to present AI ethics as a perfect foil, but to interrogate how AI safety, an epistemic community that claims to prevent human catastrophe, compares to AI ethics, a field that has a history of contending with the societal harms of AI. 

\section{Methodology: What Counts as Safety in Published Research}
\label{sec:method}

\begin{table*}[t]
\centering
\small
\caption{Of 69,344 papers collected, 54 without a usable abstract are excluded, leaving 69,290: machine learning 52,977 (76.5\%), natural language processing 14,605 (21.1\%), and ethics and society 1,708 (2.5\%). A dash (—) indicates the conference was not held that year.}
\label{tab:corpus}
\begin{tabular*}{\textwidth}{@{\extracolsep{\fill}} ll rrrrrr r @{}}
\toprule
& & \multicolumn{6}{c}{\textbf{Year}} & \\
\cmidrule(lr){3-8}
\textbf{Area} & \textbf{Venue} & 2020 & 2021 & 2022 & 2023 & 2024 & 2025 & \textbf{Total} \\
\midrule
\multirow{4}{2.5cm}{Machine Learning}
 & NeurIPS & 1,898 & 2,334 & 2,671 & 3,218 & 4,034 & 5,286 & 19,441 \\
 & ICML    & 1,084 & 1,183 & 1,233 & 1,828 & 2,610 & 3,330 & 11,268 \\
 & ICLR    &   687 &   859 & 1,094 & 1,573 & 2,260 & 3,703 & 10,176 \\
 & AAAI    & 1,610 & 1,694 & 1,369 & 1,720 & 2,517 & 3,182 & 12,092 \\
\midrule
\multirow{3}{2.5cm}{Natural Language Processing}
 & ACL     &   778 &   708 &   700 & 1,072 &   937 & 1,698 & 5,893 \\
 & EMNLP   &   751 &   846 &   825 & 1,046 & 1,264 & 1,807 & 6,539 \\
 & NAACL   &     — &   476 &   441 &     — &   561 &   695 & 2,173 \\
\midrule
\multirow{3}{2.5cm}{Ethics \& Society}
 & FAccT   &    70 &    74 &   174 &   144 &   167 &   206 & 835 \\
 & AIES    &    68 &   109 &   105 &    97 &   150 &   238 & 767 \\
 & EAAMO   &     — &    19 &    21 &    38 &    13 &    15 & 106 \\
\midrule
\multicolumn{2}{@{}l}{\textbf{Total}} & 6,946 & 8,302 & 8,633 & 10,736 & 14,513 & 20,160 & \textbf{69,290} \\
\bottomrule
\end{tabular*}
\end{table*}

Frontier safety frameworks define what counts as safety in policy (\S~\ref{sec:frontier-framweworks}), and peer-reviewed published research is usually where that definition is put into practice. We examine the three parts of our question introduced in \S~\ref{sec:intro} across six years of accepted papers at ten major venues. For trajectory, we measure the share of safety-, ethics-, and race-focused research by year and venue; for practice, the subdomains, methods, and kinds of evidence of safety research, compared with ethics; and for concepts, whether abstracts mention harm to people, engage with race or name Black communities, and which vocabulary, safety or ethics, they use for that harm.


\subsection{Corpus: Ten Venues, Three Research Areas}
\label{sec:corpus}

We collected every accepted main-track paper from ten venues in three research areas (2020–2025)\footnote{Papers from 2026 were collected but are excluded because several 2026 proceedings were not yet published at the time this research was produced.}: \textbf{Machine Learning (ML):} Neural Information Processing Systems (NeurIPS), International Conference on Machine Learning (ICML), International Conference on Learning Representations (ICLR) and the Conference on Artificial Intelligence (AAAI). \textbf{Natural Language Processing (NLP):} Association for Computational Linguistics (ACL), Chapter of the Nations of the Americas da Association for Computational Linguistics (NAACL), Empirical Methods in Natural Language Processing (EMNLP). \textbf{Ethics \& Society:} ACM Conference on AI, Ethics, and Society (AIES), ACM Conference on Fairness, Accountability, and Transparency (FAccT), ACM Equity and Access in Algorithms, Mechanisms, and Optimization (EAAMO). 

We call the ML and NLP venues \textit{mainstream} venues because they host most peer-reviewed technical safety work; we also include the main ethics venues, since they are the primary home of fairness, accountability, and auditing research. Including both lets us observe how these two communities cover risks caused by AI systems through the lens of ethics, safety, or both. We collected titles and abstracts from the official proceedings and keep a paper only if it has a usable abstract: of 69,344 papers collected, \textbf{69,290} enter the analysis, 76.5\% from ML venues, 21.1\% from NLP venues, and 2.5\% from ethics venues (Table~\ref{tab:corpus}). Details about the corpus collection are in Appendix~\ref{app:corpus}.





\subsection{Definitions are a Design Choice}
\label{sec:definitions}

Our analysis relies on language model (LM) classification, and our models do not learn from the data what counts as safety: we provide definitions, and those definitions are our main design decision. Given a paper's title and abstract, we assign exactly one of the following major classes as the main focus:

\begin{itemize}
    \item \textbf{AI ethics}: research on harms AI systems cause to people and groups, such as discrimination, privacy loss, unaccountable decisions, and the unjust distribution of burdens \cite{shelby2023harmtaxonomy, weidinger22_taxonomy_risks}.
    \item \textbf{AI safety}: research on aligning, interpreting, or securing highly capable systems against misalignment, misuse, loss of control, CBRN uplift, and catastrophic risk.
    \item \textbf{Reliability and eXplainable AI (XAI)}: robustness, uncertainty, and explanation for narrow or deployed models.
    \item \textbf{General ML}: papers with no safety, ethics, or reliability focus.
\end{itemize}

For safety, we adopt the definition that has become institutionally dominant in frontier-developer safety frameworks \cite{metr_2025_common_safety_policies, buhl2025emergingpracticesfrontierai} and that the field's own research agendas use \cite{bengio2025singaporeconsensusglobalai, anthropic2025_safety_directions}. It is narrower than the \textit{International AI Safety Report}, which groups risks into malicious use, malfunctions, and systemic risks \cite{bengio2026internationalaisafetyreport}. 

We deliberately adopt the narrower definition, because it governs which risks trigger action in corporate frameworks (\S~\ref{sec:frontier-framweworks}). We also keep generic reliability gains from being counted as safety and place them in a different class (XAI). Full definitions and the scope rules for borderline cases, such as social bias in LLMs, interpretability, privacy, and adversarial robustness, are in Appendix ~\ref{app:class-def-score-rules}.

\subsubsection{Taxonomies} Each class has its own subdomain taxonomy (Table ~\ref{tab:taxonomy}). The 20 ethics subdomains are built on \citet{shelby2023harmtaxonomy} and \citet{weidinger22_taxonomy_risks}, and 19 safety subdomains in seven groups are adapted from published safety research agendas \cite{bengio2025singaporeconsensusglobalai, anthropic2025_safety_directions}. 

\begin{table*}[ht]
\centering
\footnotesize
\caption{Subdomain taxonomies and confirmed papers per primary subdomain. Bold rows give group totals and their share of the class. Interpretability, red-teaming and alignment training account for 72\% of safety papers; control, scheming and deception, dangerous-capability evaluations and biorisk together account for 28. Definitions in Appendix~\ref{app:harm-subdomain}.}
\label{tab:taxonomy}
\begin{minipage}[t]{0.49\textwidth}
\centering
\begin{tabular}{@{}l r r@{}}
\toprule
\multicolumn{3}{@{}l}{\textbf{Ethics Class}, $n = 4{,}226$} \\
\midrule
 & Papers & \% \\
\midrule
\textbf{Fairness \& Discrimination} & \textbf{1,778} & \textbf{42.1} \\
\quad Allocative fairness & 883 & \\
\quad Representational harms \& social bias & 690 & \\
\quad Quality-of-service disparities & 205 & \\
\midrule
\textbf{Privacy \& Surveillance} & \textbf{976} & \textbf{23.1} \\
\quad Privacy & 949 & \\
\quad Surveillance \& policing & 27 & \\
\midrule
\textbf{Information, Society \& Values} & \textbf{723} & \textbf{17.1} \\
\quad Normative \& philosophical foundations & 313 & \\
\quad Copyright, consent \& provenance & 190 & \\
\quad Misinformation \& public discourse & 187 & \\
\quad Environmental \& societal impact & 33 & \\
\midrule
\textbf{Accountability \& Transparency} & \textbf{194} & \textbf{4.6} \\
\quad Documentation \& transparency & 93 & \\
\quad Contestability \& recourse & 70 & \\
\quad Explanations for affected people & 31 & \\
\midrule
\textbf{Participation \& Power} & \textbf{194} & \textbf{4.6} \\
\quad Participatory \& community-centered design & 94 & \\
\quad Labor \& data work & 58 & \\
\quad Global South \& decolonial perspectives & 42 & \\
\midrule
\textbf{Governance \& Law} & \textbf{183} & \textbf{4.3} \\
\quad Regulation \& standards & 140 & \\
\quad Public-sector AI & 25 & \\
\quad Anti-discrimination \& rights law & 18 & \\
\midrule
\textbf{Auditing} & \textbf{178} & \textbf{4.2} \\
\quad Audit methods, tools \& access & 111 & \\
\quad Algorithmic \& bias audits & 67 & \\
\bottomrule
\end{tabular}
\end{minipage}\hfill
\begin{minipage}[t]{0.49\textwidth}
\centering
\begin{tabular}{@{}l r r@{}}
\toprule
\multicolumn{3}{@{}l}{\textbf{Safety Class}, $n = 2{,}182$} \\
\midrule
 & Papers & \% \\
\midrule
\textbf{Interpretability \& Understanding} & \textbf{779} & \textbf{35.7} \\
\quad Interpretability & 761 & \\
\quad Monitoring & 18 & \\
\midrule
\textbf{Robustness, Defense \& Systemic Control} & \textbf{549} & \textbf{25.2} \\
\quad Red-teaming & 422 & \\
\quad Adversarial robustness & 127 & \\
\midrule
\textbf{Capability Control \& Unlearning} & \textbf{398} & \textbf{18.2} \\
\quad Alignment training & 398 & \\
\midrule
\textbf{Threat Modeling, Elicitation \& Evals.} & \textbf{259} & \textbf{11.9} \\
\quad Safeguards & 124 & \\
\quad Harm \& trustworthiness evaluations & 91 & \\
\quad Scheming \& deception & 14 & \\
\quad Auditing \& third-party evaluation & 13 & \\
\quad Dangerous-capability evaluations & 11 & \\
\quad Model organisms & 3 & \\
\quad Control & 2 & \\
\quad Biorisk & 1 & \\
\midrule
\textbf{Scalable Oversight \& Value Learning} & \textbf{132} & \textbf{6.0} \\
\quad Scalable oversight & 113 & \\
\quad Multi-agent safety & 19 & \\
\midrule
\textbf{Agent Foundations \& Alignment Theory} & \textbf{37} & \textbf{1.7} \\
\quad Agent foundations & 37 & \\
\midrule
\textbf{Technical Governance \& Policy} & \textbf{28} & \textbf{1.3} \\
\quad Policy \& governance & 21 & \\
\quad Strategy \& forecasting & 6 & \\
\quad AI welfare & 1 & \\
\bottomrule
\end{tabular}
\end{minipage}
\end{table*}

\subsubsection{Boundaries for classifying papers addressing race and racial harms}
\label{sec:robustness}
Our definitions send social bias in LLMs to ethics unless a paper studies it mainly as a frontier-safety problem, so “race research is classified as ethics” is partly true by construction, and we do not rest the argument on it. This is intended: our question is where research falls under the definition that now governs what counts as safety, not whether race “really” belongs to safety. Because the choice could produce some findings by construction, we test our expectations with three measures that do not depend on the class label: how often safety papers engage with race at all, the vocabulary each abstract uses for the racial harm it studies, and where race-focused work is published.


\subsection{Two-stage classification}
\label{sec:classification}

We classify each paper by title and abstract with open-weight models, we use two models from the DeepSeek family, accessed through the OpenRouter API: \texttt{DeepSeek V4.1 Flash}\footnote{https://openrouter.ai/deepseek/deepseek-v4.1-flash} for Stage 1 and \texttt{DeepSeek V4.1 Flash 1017}\footnote{https://openrouter.ai/deepseek/deepseek-v4.1-flash-1017} for Stage 2. We use a model rather than keyword matching because the distinction depends on context: for example, “alignment” can mean image registration or value alignment, and “bias” can be statistical or social. A low-cost first stage screens all papers; a second stage reads only the candidates and codes them in detail. Prompts, model settings, and costs are in Appendix~\ref{app:model-prompts}; validation details are in Appendix~\ref{app:validation}.


\subsubsection{Stage 1: screening for safety and ethics papers} The model assigns each paper to one of the four classes based on its main focus. The screen is tuned for recall: when unsure, the model keeps a paper as ethics or safety, so that a wrongly kept candidate can be removed in Stage 2, while a wrongly dropped one would be lost (Appendix~\ref{app:prompt-screen}). Of 69,290 papers with a usable abstract, \textbf{8,374 (12.1\%)} were screened in: \textbf{5,202 ethics (7.5\%)} and \textbf{3,172 safety (4.6\%)}.

\subsubsection{Stage 2: confirmation and coding} We coded each kept paper again with a longer prompt that includes worked boundary cases for differential privacy, fairness algorithms, unlearning, attacks on language models versus narrow models, and interpretability. The class from Stage 1 is fixed; the model can record disagreement in a separate field. Stage 2 records:

\begin{itemize}
    \item \textbf{Motivation gate}: how central the class’s concern is to the paper (1 = side note, 2 = co-equal goal, 3 = primary goal). Papers scoring 1 leave the study; we call papers scoring 2–3 \textit{confirmed}.
    \item \textbf{Research practice}: a primary subdomain from the class's taxonomy (Table~\ref{tab:taxonomy}), and multi-label method and evidence types (Appendix~\ref{app:prompt-subdomain}).
    \item \textbf{Social harm}: records whether the abstract mentions any harm AI causes to people or social groups, even in one sentence; generic “harmful content” or catastrophic risk alone does not count; records the vocabulary used for that harm, independently of class: safety, ethics, both, or neither.
    \item \textbf{Race topic coverage}: every confirmed paper is coded for race, whatever its class. The model first copies the phrase from the abstract that shows how the paper treats race, then assigns \texttt{race\_focus}: central, substantial, incidental (race only as a dataset attribute) or none. Appendix~\ref{app:prompt-subdomain} describes further race fields, including the communities named and the vocabulary used for the racial harm.
\end{itemize}

We treat race as a social and structural category rather than a fixed attribute \cite{Hanna2020_race_fairness}, and we code only what abstracts name. Naming a community is therefore a property of how a paper presents its work, not of who it ultimately affects. Stage 2 confirmed 4,226 of 5,202 ethics papers (81.4\%) and 2,182 of 3,172 safety papers (68.8\%): 6,408 confirmed papers, 9.2\% of the corpus.

\subsection{Analysis \& Validation}
\label{sec:validation}

We report shares of all accepted papers, per year and per venue family, so that growth is not confused with the growth of the conferences. Within the confirmed set, we report counts with Wilson 95\% confidence intervals, which cover sampling error but not classification error; where cells are small, we report counts rather than percentages. Because a language model's labels are only as useful as they are trustworthy, we test them in three ways: against a second model, against human judgment where the two models disagree, and against an independent classification of the same venues (Appendix~\ref{app:validation}).

\begin{table}[h]
\centering
  \small
  \caption{Agreement (Agr.) between the study model (\texttt{DeepSeek V4.1 Flash}, \texttt{DeepSeek V4.1 Flash 1017}) and a second model (\texttt{Gemma 4 26B A4B}) per classification stage (stg.), weighted by the corpus, except for the 254 race-focused papers (a census).}
  \label{tab:agreement}
  \setlength{\tabcolsep}{3pt}
  \begin{tabular}{@{}lp{0.40\linewidth}cc@{}}
    \toprule
    Stg. & Label & Agr. & Cohen's $\kappa$ (95\% CI) \\
    \midrule
    1 & Keep (ethics or safety) vs.\ drop & 0.98 & 0.89 (0.81--0.94) \\
    2 & Confirmed (motivation $\geq$ 2)  & 0.85 & 0.46 (0.30--0.61) \\
    2 & Mentions social harm             & 0.85 & 0.70 (0.60--0.80) \\
    2 & Names Black communities          & 1.00 & 0.91 (0.86--0.95) \\
    2 & Race-focused (all 254)           & 0.96 & --                \\
    \bottomrule
  \end{tabular}
\end{table}

\paragraph{\textbf{A second model.}} We re-ran both stages on stratified samples with an open-weight model from a different family, \texttt{Gemma 4 26B A4B}, using the same prompts, title and abstract input, and reasoning budget, including all 254 race-focused papers, and weight agreement back to the corpus (Table~\ref{tab:agreement}; samples in Appendix~\ref{app:validation}). We report agreement and Cohen's $\kappa$, which corrects for chance (Appendix~\ref{app:agreement}). The race and social-harm labels, on which our main findings rest, reach $\kappa$ of 0.70–0.91. The correction matters when one label dominates: the motivation gate has 85\% agreement but $\kappa = 0.46$, because 77\% of screened papers pass the gate, so much of that agreement would occur by chance; Gwet's AC1, which is robust to this, is 0.79. (Appendix~\ref{app:agreement}) Where the two models disagree on the gate, human review sides with the study model in 29 of 34 cases.

\textbf{Human judgment where the models disagree.} Where two models disagree, one of them is wrong. One author labeled every disagreement in the two samples from the title and abstract alone, using the prompt definitions: 131 labels on 124 papers. The human label matches the study labels in 96 of 131 cases (73\%; 95\% CI 65–80\%) and, on the three race fields, in 30 of 35 (86\%; 95\% CI 71–94\%). The screening errors tend toward inclusion, as intended, and Stage 2 removes such borderline papers. All four errors in naming Black communities are papers the study missed, so the count of papers naming Black communities is a lower bound.

\textbf{An independent classification.} Compared with \citet{Roytburg2026}, who classified papers from overlapping venues, our ethics labels agree closely, while our safety class is narrower because theirs includes robustness and explainability. This confirms that the scope of safety is a definitional choice (\S~\ref{sec:definitions}), and that our safety counts describe frontier safety specifically (see Appendix~\ref{app:mtg} for more comparison values).

\section{Findings}
\label{sec:findings}

\subsection{Trajectory: where safety and race research are published} 
\label{sec:trajectory}

\begin{figure*}[!t]
  \centering
  \begin{minipage}[t]{0.448\textwidth}
    \centering
    \includegraphics[width=\linewidth]{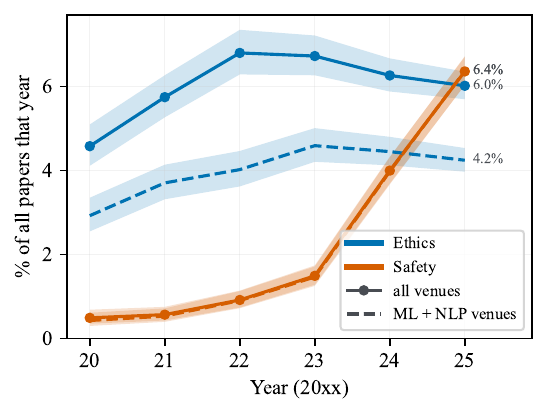}
    \caption{Classified AI safety and AI ethics papers as a share of all accepted papers, 2020–2025. Solid lines: all ten venues; dashed lines: the seven ML and NLP venues. Safety rose from 0.5\% of papers (34) in 2020 to 6.4\% (1,282) in 2025 and overtook ethics (6.0\%). At ML and NLP venues alone, ethics fell to 4.2\%.}
    \Description{}
    \label{fig:share_over_time}
  \end{minipage}\hfill
  \begin{minipage}[t]{0.49\textwidth}
    \centering
    \includegraphics[width=\linewidth]{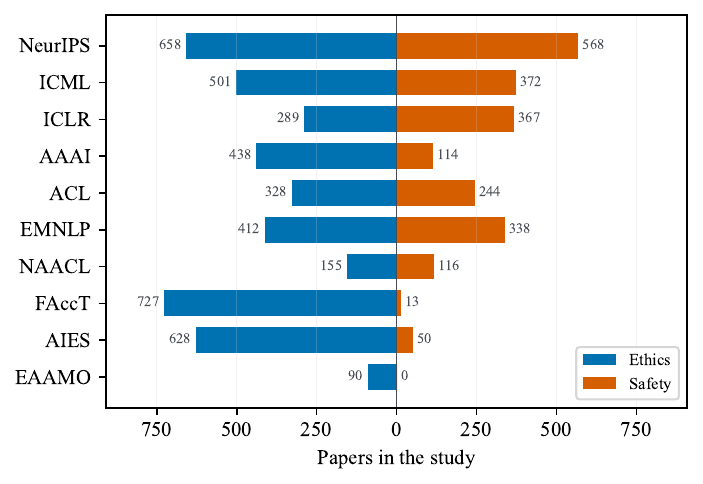}
    \caption{Confirmed ethics (left) and safety (right) papers by venue, 2020–2025 (n = 4,226 ethics; 2,182 safety). Safety is published almost entirely at ML and NLP venues (97.1\%); the three ethics venues publish 63 safety papers. Ethics appears at every venue.}
    \Description{Horizontal butterfly bar chart showing confirmed ethics (left) and safety (right) papers by venue from 2020 to 2025. Ethics papers are distributed across all venues, while safety papers are concentrated in ML and NLP venues.}
    \label{fig:venues_butterfly}
  \end{minipage}
\vskip -0.12in
\end{figure*}

Confirmed AI safety research grew from 34 papers (0.5\% of accepted papers) in 2020 to 1,282 (6.4\%) in 2025, a 38-fold increase in papers and roughly a 13-fold increase in share (Fig.~\ref{fig:share_over_time}). Over the same period, AI ethics remained between 5\% and 7\% of accepted papers (318 to 1,213). The inflection falls between 2023 and 2024, when safety rose from 160 papers (1.5\%) to 580 (4.0\%), consistent with accounts that place the institutional rise of safety in 2023 \cite{ahmed2024field, lazar2023safetyterms}.

In 2025, safety overtook ethics: 1,282 papers against 1,213 overall, and 1,255 against 836 at mainstream venues (6.4\% versus 4.2\% of mainstream papers). The confirmed series follows the same shape as the screen series, about 1.5–2 percentage points lower, so the trend is not an artifact of the screen's recall setting (Fig.~\ref{fig:class_share_per_year}, Appendix~\ref{app:screen-confirmed}), and safety also became more committed as it grew. 


The two fields publish in different places (Fig.~\ref{fig:venues_butterfly}). Of confirmed safety papers, 97.1\% appear at mainstream venues; only 63 appear at the three ethics venues. Ethics, by contrast, is published everywhere: 65.8\% of ethics papers appear at mainstream venues, and the share appearing at ethics venues fell from 40\% (2020–22) to 32\% (2023–25). Ethics therefore did not retreat into its own venues. It grew slowly while safety grew quickly in the same venues.

\begin{figure*}[t!]
\centering
\includegraphics[width=\textwidth]{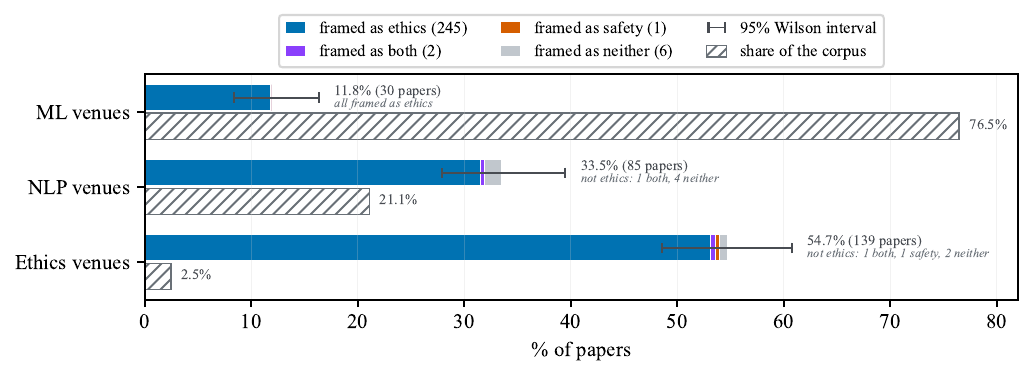}
\caption{Race-focused papers (n = 254) by venue family, compared with each family's share of the corpus (hatched). Ethics venues publish 2.5\% of the corpus but 54.7\% of race-focused papers (95\% Wilson interval 48.6–60.7\%); ML venues publish 76.5\% of the corpus but 11.8\% of race-focused papers. Colors show the vocabulary each abstract uses for the racial harm: 245 papers frame it as ethics, 3 as safety or both, and 6 as neither.}
\label{fig:race_where_framing_single}
\Description{Bar plot comparing the proportion of race-focused papers by venue family with each family's share of the corpus, using color to indicate vocabulary framing.}
\end{figure*}

Research on racial harm did not follow safety into the mainstream. Race-focused ethics papers grew from 17 in 2020 to 69 in 2025, but stayed at 5–7\% of ethics research every year; papers naming Black or African-diaspora communities stayed between 1.1\% and 2.3\% (7 in 2020, 24 in 2025). Of the 254 race-focused papers, 139 (55\%) appear at FAccT, AIES, and EAAMO, venues that publish 2.5\% of the corpus (Fig.~\ref{fig:race_where_framing_single}). NLP venues publish 85. The four ML venues, where most safety research is published, carry 30 race-focused papers among 3,307 confirmed safety and ethics papers.

\textbf{Expectation 1 holds.} As safety grew 13-fold in share, research on racial harm grew only with ethics, stayed at the same share of it, and remained concentrated in the three venues that safety research barely enters. The contradicting result, race-focused research rising at mainstream venues alongside safety, does not appear.


\subsection{Practice: a narrow, model-centered agenda} 
\label{sec:practice}

Three subdomains account for 72\% of safety papers: interpretability (761 papers, 34.9\%), red-teaming (422, 19.3\%), and alignment training (398, 18.2\%) (Table~\ref{tab:taxonomy}). The catastrophic risks that motivate the field's public case are almost absent from its peer-reviewed output. Control, scheming and deception, dangerous-capability evaluations, and biorisk together account for 28 papers (1.3\%).

Over time, safety shifted from understanding models to attacking and constraining language models, while ethics stayed stable (Appendix~\ref{app:groups-over-time}). 


\begin{table}[t]
\centering
\small
\caption{Method and evidence types, share of confirmed papers (multi-label, so shares do not sum to 100\%). Safety relies on benchmarks and model internals; evidence from people, deployed systems, and audits is rare.}
\label{tab:methods_evidence}
\begin{tabular}{@{}p{0.35\linewidth}rr@{}}
\toprule
 & Safety (\%) & Ethics (\%) \\
 & $n = 2{,}182$ & $n = 4{,}226$ \\
\midrule
Automated metrics & 93.8 & 68.5 \\
Model internals & 37.6 & 2.2 \\
Attacks & 21.4 & 3.5 \\
Human studies & 1.1 & 10.5 \\
Participant data & 1.1 & 9.8 \\
Deployed systems & 0.8 & 4.8 \\
Audits & 0.4 & 4.9 \\
\bottomrule
\end{tabular}
\end{table}

Safety research is also model-centered in its methods and evidence (Table~\ref{tab:methods_evidence}).  Almost all safety papers rely on automated metrics (93.8\%, against 68.5\% of ethics papers); 37.6\% analyze model internals and 21.4\% design attacks. Evidence drawn from people and institutions is nearly absent: about 1\% of safety papers study people, 0.8\% draw on deployed systems, and 0.4\% audit a system. The research that documents harm to Black communities is built from exactly this missing evidence, of the 77 papers that name Black or African-diaspora communities, 21\% draw on participants, 19\% on deployed systems, and 14\% on audits.

\textbf{Expectation 2 holds.} The evidence safety accepted is produced by measuring models against benchmarks, not by studying the people and institutions where models are deployed, which is how the majority of research in our study has been documenting harm to Black communities. Safety uses participant, deployment, and audit evidence at a small fraction of the rate in ethics research.

\subsection{Concepts: whose harm counts} 
\label{sec:concepts}
 
\textbf{Harm to people:} About one in nine safety papers mentions harm to people or social groups, even in a single sentence: 246 of 2,182 (11.3\%; 95\% CI 10.0–12.7\%), against 3,286 of 4,226 ethics papers (77.8\%; 76.5–79.0\%). The share did not grow as the field grew more than thirtyfold: it was 10.6\% in 2020–22 and 11.3\% in 2023–25. The largest subdomains are the least people-facing (Appendix~\ref{app:harm-subdomain}).

\textbf{Two vocabularies:} When safety papers name what goes wrong, they name what the model does (\textit{harmful, attacks, outputs, jailbreak}); ethics papers name who is affected (\textit{bias, fairness, groups, social}). Only 8.8\% of safety papers use ethics vocabulary at all, and 37\% frame no harm at all.

\textbf{Race in safety research:} Of 2,182 confirmed safety papers, 8 mention race at all, 6 of them only incidentally, and none name Black or African-diaspora communities (0 of 2,182; 95\% CI 0–0.18\%; Appendix~\ref{app:race-venues}). The absence is clearest where race would be most expected: \textit{harm and trustworthiness evaluations} is the one safety subdomain whose definition includes biased and unfair outputs, and none of its 91 papers makes race a central or substantial topic in the research paper.

\textbf{Racial harm is rarely described as a safety problem:} Across all 254 classified as race-focused papers, the abstract describes the racial harm in safety vocabulary, or in both vocabularies, in only 3 cases (1.2\%; 95\% CI 0.4–3.4\%) (Fig.~\ref{fig:race_where_framing_single}, colors). This holds even for research on safety's own interventions. Papers on detoxification (NAACL 2021), toxicity and hate-speech classifiers (AAAI 2020, EAAMO 2021, NAACL 2022, FAccT 2022), and alignment training (ACL 2025) find that these interventions harm racialized users, especially speakers of African American English, yet they are framed and published as ethics. The one race-focused paper caught by our classification procedure frames its argument as safety and names Black communities, argues from outside the field for community-based safety, and appears at an ethics venue \cite{Sandoval2025_rethinking_safety}.

\textbf{Who is named:} Most of the classified race-focused papers do not say which communities they study: 137 refer only to “racial or ethnic minorities.” Black or African-diaspora communities are named in 77 papers, all ethics: 1.8\% of ethics research and 0.1\% of the corpus, a lower bound (\S~\ref{sec:validation}; Appendix~\ref{app:race-details}).
 
\textbf{Expectation 3 holds.} Under the definition of safety that now governs the field, harm to Black communities is not seen as a safety topic. According to our study, no safety paper names Black communities, and racial harm is described in safety terms in only 3 of 254 cases, even when the harm is caused by safety’s own interventions. This result does not depend on our class labels; the racial-harm vocabulary is coded independently of class (\S~\ref{sec:robustness}).

\section{Discussion of the Results}
\label{sec:discussion}
\label{sec:wynter-reading}

Our findings show that AI safety does not address anti-Blackness. The divergence in what is viewed as harm influences how AI safety research, AI ethics, and research that focuses on anti-Blackness and racial harms frame it; In safety research, harm is defined around the model: abstracts name what the model does (attacks, outputs, jailbreaks) rather than who is affected; 37\% frame no harm, and not one of 2,182 safety papers names Black communities. Participant studies, deployed cases, and audits documenting harm to Black communities make up about 1\% or less of the evidence in AI safety research. Surprisingly, despite the threat of catastrophic harm being central to the beliefs of the AI safety epistemic community and the frontier frameworks that cement them into institutional practice (\S~\ref{sec:ai-safety-roots}), we found that addressing catastrophic harms was significantly low in AI safety research. Research on catastrophic risk counts for 28 papers (1.3\%) of its output. 

Research on harm to Black communities roots its work via evaluating and framing harm based on its impacts on people, the consequences of deployed systems, and assessing auditing procedures. Documented, cumulative harms to Black communities in health, hiring, speech recognition, and language technologies are mainly coded as bias and published as ethics. The line between catastrophe and bias is therefore not drawn by severity; it follows whose harm is at stake.

A field that claims to protect humanity while excluding the harms borne by Black communities narrows both its relevance and ability to protect all people, and limits the scientific methods and analysis it pursues. \citeauthor{Wynter2003}’s critique explains how a field that claims to want to save us all rarely addresses the public in its research; the share of safety papers mentioning harm to people stayed near 11\% while the field grew more than thirtyfold (\S~\ref{sec:concepts}). Safety’s “us” is the overrepresented figure of Man, and its catastrophes are the ones that would reach that figure. The lineage that \citet{GebruTorres2024_TESCREAL} trace from first-wave eugenics to the TESCREAL bundle shows that this is more than a metaphor: the conception of intelligence at the root of the safety movement was built on ranking human worth, the same ranking that coloniality used to place Black and Indigenous peoples outside the human \cite{Wynter2003}. 
As safety overtakes ethics in the mainstream venues (\S~\ref{sec:trajectory}), and as its frameworks decide which evaluations get built (\S~\ref{sec:frontier-framweworks}), this narrow scope risks becoming the default for AI evaluation as a whole.

\section{Conclusion and Future Work}
\label{sec:conclusion}

Investigating the state of addressing anti-Blackness in the age of AI safety, our findings present a bleak picture. AI safety has come to dominate AI politics, institutional frameworks, and even risen in prominence in research. Yet, through \citeauthor{Wynter2003}'s conception of Man, we can see that AI safety does not intend to save us all from catastrophic harm, nor can it calculate what will be catastrophic to the Human. Our research corroborates Wynter's conception of Man due to the complete absence of attempts to address anti-Blackness and racial harm in AI safety literature.  This is a trend that researchers, engineers, and policymakers must resist. It is enticing to abandon addressing anti-Blackness and racial harms at a time when research and policy even associated with Diversity, Equity, and Inclusion (DEI) are being attacked. However,  \citet{Wynter2003} shows that the deeply catastrophic and near-totalizing effect of anti-Blackness does not allow us to secure the well-being of the Human if we do not contend with and ultimately \textit{unsettle} the anti-Black, racial logics that have constructed Modernity. There is no just technological future if we do not confront anti-Blackness directly.  


In future research, we aim to further explore what constitutes robust assessment to address anti-Blackness in AI systems, development, and evaluation, and how to foster interdisciplinarity in the space. We believe it is imperative that an approach to AI evaluation that furthers anti-Blackness is not entrenched in AI research, development, evaluation, and oversight; we must work toward an AI assessment, evaluation, and regulatory system accountable to Black communities. This would orient research and governance toward, as Wynter writes, “securing the well-being, and therefore the full cognitive and behavioral autonomy of the human species itself/ourselves” \cite{Wynter2003}.

\subsection{Limitations}
 
We code titles and abstracts only, so race discussed only in a paper’s body is not counted; our figures describe what each field chooses to put forward. Race is coded only in the ethics and safety classes, so race-focused work that the screen placed in reliability or general ML is not counted. We count Black communities when the abstract names them; human review found that the study labels miss abstracts that imply them (for example, “dark-skinned patients”), so the 77 papers naming Black communities are a lower bound. One author labeled the model disagreements from titles and abstracts, using the prompt definitions, so we cannot report agreement between annotators, and papers on which both models agree and are both wrong are not covered. As part of our work extension, we plan to re-code a subsample from full texts, added journals and workshops at length.


The motivation gate is the weakest label ($\kappa = 0.46$; AC1 = 0.79), so confirmed counts carry more uncertainty than the race and harm labels. Finally, our corpus covers ten venues and main tracks only, excluding workshops, journals, preprints, and industry reports, and our definition of safety is narrower than some in use (Appendix~\ref{app:mtg}). We reduce the dependence on our own definitions with three label-independent measures (\S~\ref{sec:robustness}) and release the codebook and data so that others can re-run the analysis.


\begin{acks}

\end{acks}

\section*{Generative AI Usage Statement}
Language models were core to our method, as described in \S~\ref{sec:classification} and Appendix~\ref{app:model-prompts}. \texttt{DeepSeek V4.1 Flash} and \texttt{DeepSeek V4.1 Flash 1017}, accessed through the OpenRouter API, classified papers from their titles and abstracts using a codebook written by the authors; \texttt{Gemma 4 26B A4B} re-coded stratified samples for validation, and one author reviewed every disagreement between the two models. In preparing the manuscript, \texttt{Claude Opus 5.5} was used to find related work, proofread, ideate the structure of sections, edit for grammar, revise text for the background and discussion from the authors' existing draft, and assist in reviewing code and experimental results. The authors reviewed, revised, and verified all text, numbers, and citations, and take full responsibility for the content.
 
\section*{Ethical Considerations and Adverse Impacts}
Our study analyzes titles and abstracts of publicly available, peer-reviewed papers and it involves no human participants and no personal data beyond author names already published in proceedings. Classifying other researchers’ work risks mischaracterization, so we report validation results, treat our count of papers naming Black communities as a lower bound, and release the codebook and labels so classifications can be checked and contested \footnote{Our code and datasets will be made available upon acceptance.} We describe individual papers only where they illustrate a finding, and we critique fields and definitions rather than individual researchers or works. A further risk is that our findings are read as a case for absorbing race research into safety on safety’s existing terms. A positionality statement is withheld for anonymity and will be added to the camera-ready version.


\bibliographystyle{ACM-Reference-Format}
\bibliography{references}

\onecolumn
\appendix
\section{Corpus Collection}
\label{app:corpus}

Our collection process was performed by scraping titles and abstracts from official proceedings online pages (PMLR\footnote{https://proceedings.mlr.press/}, NeurIPS proceedings\footnote{https://papers.nips.cc/}, OpenReview, and the AAAI OJS\footnote{https://ojs.aaai.org/} issue pages), and in some cases where proceedings are not automated script-readable, we used OpenAlex\footnote{https://openalex.org/} and selected papers by each venue's DOI prefix. We exclude workshop papers, demos and students volumes, non-archival abstracts, findings papers, and datasets-and-benchmarks tracks. We had some small-number cases where we failed to collect the abstract from the source, and we treat these as missing; a paper enters our analysis only if it has a usable, non-empty abstract of at least 50 characters. The collected corpus contains $69,344$ papers; we drop 54 papers that do not meet our criterion (24 NAACL, 15 EAAMO, 7 ACL, 7 EMNLP, 1 FAccT), leaving \textbf{69,290} for analysis (Table~\ref{tab:corpus}). 




\section{Class Definitions and Scope Rules}
\label{app:class-def-score-rules}
 
\begin{itemize}
    \item \textbf{AI Ethics \& Fairness.} AI ethics is research on understanding, measuring, preventing, or governing harms that AI systems cause to people and groups: discrimination, bias, and exclusion; privacy loss and surveillance; unaccountable decisions; and the unjust distribution of benefits and burdens \cite{shelby2023harmtaxonomy, weidinger22_taxonomy_risks}, usually in systems deployed or deployable today.
 
    \item \textbf{AI Safety: Frontier \& Misalignment.} Research on understanding, interpreting, aligning, or securing highly capable AI systems against misalignment, misuse, and loss of control; chemical, biological, radiological, and nuclear (CBRN) uplift; and autonomous capabilities and catastrophic risk. We adopt the definition that has become institutionally dominant in frontier-developer safety frameworks \cite{metr_2025_common_safety_policies, buhl2025emergingpracticesfrontierai} and in the field's own research agendas \cite{bengio2025singaporeconsensusglobalai, anthropic2025_safety_directions}.\footnote{Standards in this domain are evolving quickly; this definition reflects the dominant institutional definition at the time of writing.} It is narrower than the \textit{International AI Safety Report}, which groups risks into malicious use, malfunctions, and systemic risks \cite{bengio2026internationalaisafetyreport}; we adopt it deliberately because it governs which risks trigger action in corporate frameworks (\S~\ref{sec:frontier-framweworks}).
 
    \item \textbf{Truthfulness, Reliability \& eXplainable AI (XAI).} Truthfulness, reliability, and explainability for narrow or deployed models, including uncertainty estimation, robustness of narrow vision models, and post-hoc explanation \cite{Barredo2020_XAI}. Both fields claim this work: older technical agendas count robustness as safety \cite{hendrycks2022unsolvedproblemsmlsafety}, while explanations for affected people are an ethics concern \cite{miller2019explanation}. Keeping it in a separate class prevents generic reliability gains from being counted as safety \cite{Ren2024_Safetywashing}.
 
    \item \textbf{General ML \& Other.} Any paper with no safety, ethics, or reliability focus, such as work on performance, efficiency, or generalization; the class also covers learning theory, scientific applications, linguistic analysis, and datasets or resources.
\end{itemize}
 



\subsection{Scope rules}
\label{app:scope-rules}
Social bias in language models goes to the ethics class, unless the paper studies it mainly as a frontier-model safety problem (harmlessness benchmarks, red-teaming, guardrails). Interpretability that yields general insight into how models represent or reason belongs to safety, even on toy models; explanations aimed at affected people belong to ethics. Privacy-preserving learning is ethics, while data-extraction attacks on large language models (LLMs) are safety. Adversarial robustness is reliability for narrow vision models and safety for LLMs and agents.

\section{Validation Details}
\label{app:validation}
 
\subsection{Samples}
Both samples are drawn from 2020–2025 papers with a usable abstract, with a fixed random seed. The screen sample ($n = 400$) holds 100 random papers from each Stage 1 class. The Stage 2 sample ($n = 454$) holds every race-focused paper (254, a census) plus 100 random non-race papers from each of the ethics and safety classes; the second model receives the same fixed Stage 1 class. Because the samples are stratified, we also weight agreement back to the corpus: each sampled paper counts for the number of corpus papers in its stratum (69,290 papers for the screen; 254 race-focused, 4,940 ethics, and 3,170 safety papers for Stage 2). Models and settings for both labelers are in Table~\ref{tab:models_settings} (Appendix~\ref{app:model-prompts}).
 
\subsection{Agreement Measures}
\label{app:agreement}
We report agreement, the share of papers on which the two models give the same label, and Cohen's $\kappa$, which corrects it for chance ($\kappa = 0$ is chance level and $\kappa = 1$ perfect agreement); we read $\kappa$ on the usual scale, where 0.41–0.60 is moderate, 0.61–0.80 substantial, and above 0.80 almost perfect~\cite{landis1977measurement}. For each label, we report raw agreement, Cohen's $\kappa$, and Gwet's AC1 \cite{Gwet2008_variance_agreement}. $\kappa$ corrects agreement for chance, but it falls sharply when one label dominates, even when the two models almost always agree; AC1 is stable in that case. 
Cohen's $\kappa$ estimates chance agreement from how often each rater uses each label, so when one label dominates, expected chance agreement is high, and $\kappa$ falls even when raters rarely disagree, a known paradox~\cite{Feinstein1990}. Gwet's AC1 estimates chance agreement differently, so it stays stable when labels are imbalanced~\cite{Gwet2008_variance_agreement}. Confidence intervals are 95\% percentile intervals from 2,000 bootstrap resamples drawn within each stratum. Within the race census, the study labels are all race-focused by design, so $\kappa$ is undefined there, and we report agreement and AC1.
 
\subsection{Stage 1 Agreement}
Weighted to the corpus, the two models agree on keeping versus dropping a paper for 98\% of papers (95\% CI 96–99\%; $\kappa = 0.89$, 0.81–0.94; AC1 = 0.97) and on the four-way class for 95\% (93–97\%; $\kappa = 0.85$, 0.78–0.91; AC1 = 0.95). With the second model's labels, 12.5\% of the corpus would be screened in instead of 12.1\%.
 
\subsection{Screened and Confirmed Series}
\label{app:screen-confirmed}
\begin{figure}[ht]
\centering
\includegraphics[width=0.4\textwidth]{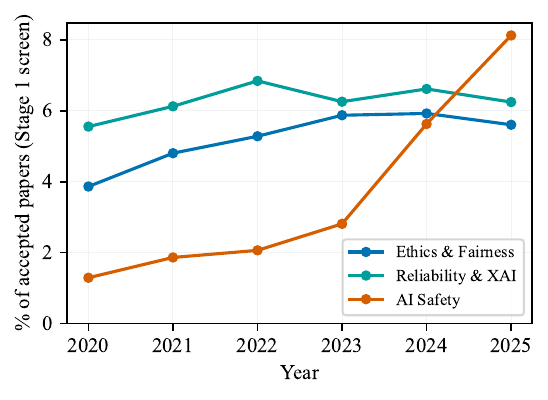}
\caption{Screened (Stage 1) and confirmed (Stage 2) safety and ethics papers as a share of all accepted papers, 2020–2025. The confirmed series follows the screened series by about 1.5–2 percentage points.}
\label{fig:class_share_per_year}
\Description{}
\end{figure}

Because the screen is tuned for recall, its counts are an upper bound. If the growth of safety research were an artifact of this setting, the screened and confirmed series would diverge over time. They do not: the confirmed series follows the screened series about 1.5–2 percentage points lower in every year, for both safety and ethics (Fig.~\ref{fig:class_share_per_year}).
 
\subsection{Stage 2 Agreement}
 
\begin{table}[ht]
\centering
\small
\caption{Stage 2 agreement between the study model and the second model, weighted to the corpus unless noted.}
\label{tab:stage2_agreement}
\begin{tabular}{@{}lrrlr@{}}
\toprule
Label & $n$ & Agreement & Cohen's $\kappa$ (95\% CI) & Gwet's AC1 \\
\midrule
Confirmed (motivation $\geq 2$) & 454 & 0.85 & 0.46 (0.30–0.61) & 0.79 \\
Mentions social harm & 454 & 0.85 & 0.70 (0.60–0.80) & 0.70 \\
Race-focused & 454 & 0.98 & 0.75 (0.56–0.98) & 0.98 \\
Names Black communities & 454 & 1.00 & 0.91 (0.86–0.95) & 1.00 \\
Subdomain group & 363 & 0.85 & 0.83 (0.76–0.90) & 0.84 \\
Race-focused (census of 254) & 254 & 0.96 & -- & 0.96 \\
Racial-harm framing (both race-focused) & 245 & 0.96 & 0.46 (0.19–0.69) & 0.95 \\
\bottomrule
\end{tabular}
\end{table}
 
\begin{figure}[ht]
\centering
\includegraphics[width=0.8\linewidth]{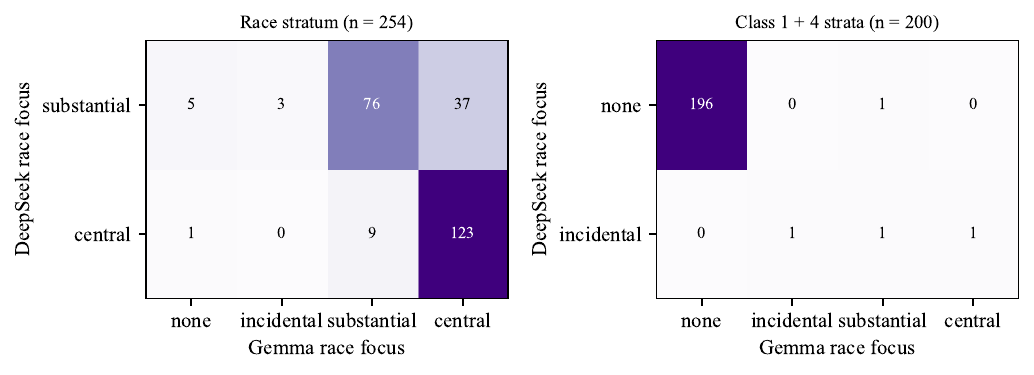}
\caption{Race focus coded by the study model and the second model on the 254 race-focused papers. Most disagreement is between substantial and central; the binary race-focused label is stable.}
\label{fig:race_focus_confusion}
\Description{Two plots of confusion matrices that show the race focus coded by the study on the 254 race-focused papers.}
\end{figure}

The second model codes 245 of the 254 race-focused papers as race-focused; most disagreement on the four race levels is between substantial and central, which is why the findings use the binary label (Fig.~\ref{fig:race_focus_confusion}). Racial-harm framing has a low $\kappa$ but high agreement because 230 of 245 papers are framed as ethics by both models.
 
\subsection{Human Annotation of the Disagreements}
We listed every paper on which the two models disagree on a label the findings rely on: 131 rows on 124 papers. One author labeled each row from the title and abstract, following a guide that restates the prompt definitions (released with the data). We state whether model labels were visible during the disagreement labeling, and that the author's background is in AI engineering, with ethics and safety knowledge in the background.
 
\begin{table}[ht]
\centering
\small
\caption{Share of model disagreements where each model's label matches the human label (Wilson 95\% intervals for the totals).}
\label{tab:human_annotation}
\begin{tabular}{@{}lrll@{}}
\toprule
Label & $n$ & Study labels & Second model \\
\midrule
Screen: keep vs.\ drop & 30 & 12 (40\%) & 18 (60\%) \\
Confirmed (motivation $\geq 2$) & 34 & 29 (85\%) & 5 (15\%) \\
Mentions social harm & 32 & 25 (78\%) & 7 (22\%) \\
Race-focused & 12 & 11 (92\%) & 1 (8\%) \\
Names Black communities & 12 & 8 (67\%) & 4 (33\%) \\
Racial-harm framing & 11 & 11 (100\%) & 0 (0\%) \\
\midrule
All labels & 131 & 96 (73\%; 65–80\%) & 35 (27\%; 20–35\%) \\
\bottomrule
\end{tabular}
\end{table}
 
The screen errs toward inclusion: all 18 screen errors are papers the study kept that the human would drop (bias in a statistical sense, security of narrow models, game theory with fairness only as a motivating example), and in none did the study drop a paper the human would keep. All four errors on naming Black communities are papers the study missed. These shares cover only papers on which the two models disagree; they are not estimates of accuracy over the corpus.

\section{Comparison with \textit{Mind the Gap!}}
\label{app:mtg}
 
\citet{Roytburg2026} classify papers from overlapping venues as safety or ethics: they first keep papers whose title or abstract matches a keyword list, then ask \texttt{Gemini 2.5 Flash} whether the main contribution is ethics or safety. We matched papers on normalized title within venue and year: 4,317 of the 5,446 papers from venues in our corpus (79.3\%).
 
\begin{table}[ht]
\centering
\small
\caption{\textit{Mind the Gap!} categories against our screening classes, on 4,317 matched papers.}
\label{tab:mtg}
\begin{tabular}{@{}lrrrrrr@{}}
\toprule
Their category & Papers & Ethics & Safety & Reliability & General ML & Ethics or safety \\
\midrule
Ethics & 2,091 & 81.1\% & 4.7\% & 4.5\% & 9.6\% & 85.8\% \\
Safety & 1,882 & 9.6\% & 27.3\% & 31.0\% & 32.1\% & 36.9\% \\
Explainability & 344 & 18.3\% & 19.2\% & 44.5\% & 18.0\% & 37.5\% \\
\bottomrule
\end{tabular}
\end{table}
 
The two classifications agree on ethics and differ on safety, for a systematic reason. Their definition follows the broad ``ML safety'' agenda \cite{hendrycks2022unsolvedproblemsmlsafety} and includes robustness, uncertainty, verification, and all explainability, which we place in the reliability class; the papers we place in general ML use safety vocabulary in another sense, such as conservative reinforcement learning. Merging safety and explainability, as they do, gives the same result (37.0\% in our ethics or safety classes). Agreement is highest at the ethics venues (FAccT 96.6\%, AIES 95.3\%) and lowest at ML venues (NeurIPS 37.3\%). Readers who prefer the broader definition can add the reliability class to safety; the released data include both.

\section{Additional Results}
\label{app:additional-results}


 
\subsection{Safety subdomains over time}
\label{app:groups-over-time}
\begin{figure}[ht]
\centering
\includegraphics[width=0.9\linewidth]{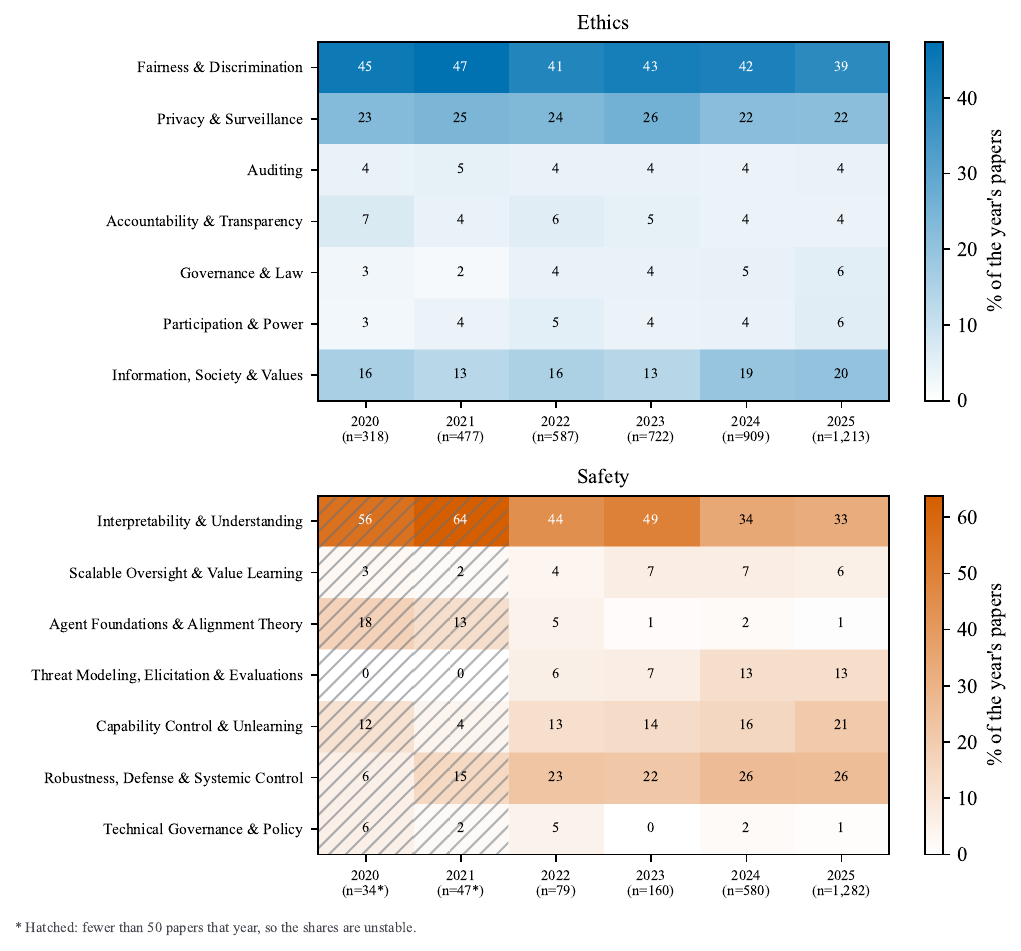}
\caption{Safety subdomain groups as a share of safety papers per year. Interpretability and understanding fell from 56\% to 33\%; robustness, defense, capability control, and threat modeling rose from 18\% to 60\%; agent foundations fell from 18\% to 1\%.}
\label{fig:groups_over_time}
\Description{Line plot showing trends in safety subdomain groups over time.}
\end{figure}

The overall share of safety research hides a change in what the field studies. In 2020, more than half of safety papers sought to understand models (interpretability and understanding); by 2025, most sought to attack or constrain them (Fig.~\ref{fig:groups_over_time}). This shift follows the move of safety research toward large language models: red-teaming, alignment training, and safeguards grew with the deployment of chat models, while agent foundations, the most theoretical group, almost disappeared. 
Interpretability and understanding fell from 55.9\% of safety papers in 2020 to 32.5\% in 2025; robustness, defense, capability control, and threat modeling together rose from 18\% to 60\%; agent foundations fell from 17.6\% to 0.6\% (Fig.~\ref{fig:groups_over_time}). Ethics, by contrast, is stable: fairness stays near 40\% of ethics papers and privacy near 22\%. What ethics means depends on the venue, however. At NeurIPS and ICML, almost half of ethics papers concern privacy, whereas at FAccT and AIES ethics is spread across governance, participation, and accountability, groups that ML venues rarely publish.
 
\subsection{Race in safety papers, by venue}
\label{app:race-venues}
\begin{figure}[ht]
\centering
\includegraphics[width=\linewidth]{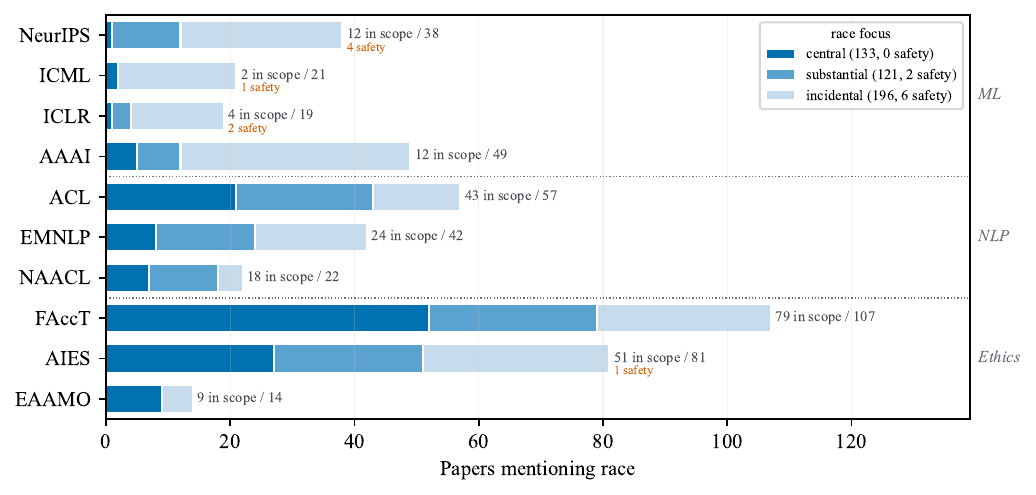}
\caption{Papers mentioning race, by venue and race focus ($n = 450$). Race-focused papers (central or substantial, $n = 254$) concentrate at FAccT and AIES; of the 8 safety papers mentioning race, none treats it as central.}
\Description{Bar chart showing the number of papers mentioning race by venue and race focus. Most race-focused papers are at FAccT and AIES; none of the 8 safety papers treats race as central.}
\label{fig:race_focus_venues}
\end{figure}

Race appears in safety research only at the margins. Of the 450 confirmed papers that mention race, 8 are safety papers: 6 mention it only as a dataset attribute and 2 treat it substantially, and none makes it central (Fig.~\ref{fig:race_focus_venues}). The venue breakdown also shows that race-focused papers at ML venues are spread thinly: NeurIPS, ICML, ICLR, and AAAI together publish 30, against 79 at FAccT alone.
 
The two race-focused safety papers do not treat racial harm as their problem: one is an interpretability method for diffusion models with a bias-analysis application (NeurIPS 2024); the other draws on Indigenous epistemologies to argue about AGI control (AIES 2020).
 
\subsection{Social harm by safety subdomain}
\label{app:harm-subdomain}
The share of safety papers that mention harm to people varies by subdomain: 2\% of interpretability papers, 5\% of scalable-oversight papers, 7\% of adversarial-robustness papers, and 8\% of safeguards papers, rising to 17\% for alignment training, 22\% for red-teaming, 26\% for harm and trustworthiness evaluations, and 46\% for auditing and third-party evaluation. The three largest subdomains (interpretability, red-teaming, and alignment training) account for 72\% of safety papers. The most frequent harm terms in safety abstracts are \textit{harmful, attacks, safety, alignment, opaque, outputs, adversarial}, and \textit{jailbreak}; in ethics abstracts, \textit{privacy, bias, fairness, unfair, groups, social}, and \textit{sensitive}. Of safety papers, 5.3\% use ethics vocabulary alone and 3.5\% use both vocabularies.
 
\subsection{Race-focused research: details}
\label{app:race-details}
\paragraph{Generative models.} About 79 race-focused papers study LLMs or other generative models, rising from 1 in 2020 to 37 in 2025; 76 of them frame the racial harm as ethics. This count comes from a keyword match on titles and harm descriptions and is approximate.
 
\paragraph{The safety-framed exception.} The one race-focused paper that uses safety vocabulary and names Black communities, \textit{Rethinking AI Safety: Provocations from the History of Community-based Practices of Road and Driver Safety} \cite{Sandoval2025_rethinking_safety}, appears at an ethics venue and argues from outside the field for what safety should include.
 
\paragraph{Papers naming Black communities.} Of the 77 papers that name Black or African-diaspora communities, 69 are race-focused. They study representational harm (48, including stereotyping in generated text and images and dialect bias against African American English), quality-of-service disparities (24), allocative harm (16), and surveillance and policing (7); 15 report intersectional findings, for example on Black women. Of the 77, 43 (56\%) appear at FAccT, AIES, and EAAMO, 29 at NLP venues, and 5 at ML venues. Their share of ethics research stays between 1.1\% and 2.3\% in every year.

\section{Models, Settings, and Prompts}
\label{app:model-prompts}
 
The full prompts, code, coded corpus, and human-labeled validation set will be available soon upon acceptance. 
This appendix summarizes the models, settings, and the information each prompt asks for. Models and settings for both stages are in Table~\ref{tab:models_settings}. Screening the whole corpus and running the detailed prompt only on kept papers costs about 81\% less than running the detailed prompt on every paper.
 
\begin{table}[ht]
\centering
\small
\caption{Models and settings for the study labels and the second model.}
\label{tab:models_settings}
\begin{tabular}{@{}lll@{}}
\toprule
 & Study labels & Second model \\
\midrule
Stage 1 model & \texttt{deepseek-v4.1-flash} & \texttt{gemma-4-26b-a4b-it} \\
Stage 2 model & \texttt{deepseek-v4.1-flash-1017} & \texttt{gemma-4-26b-a4b-it} \\
Prompts & coding repository \footnote{Our code and datasets will be made available upon acceptance.} & Same \\
Input & Title and abstract ($\leq$ 2,000 characters) & Same \\
Reasoning budget & None (Stage 1); 1,024 tokens (Stage 2) & Same \\
Output limit & 500 (Stage 1); 4,024 tokens (Stage 2) & 500 (Stage 1); 6,000 tokens (Stage 2) \\
Temperature & 1.0 & 1.0 \\
\bottomrule
\end{tabular}
\end{table}
 
\subsection{Screening prompt}
\label{app:prompt-screen}
Without reasoning and with structured output, the model assigns each paper to one of four classes based on its main focus. A paper enters the ethics or safety class only if it meets the class definition and pursues that concern at least as a co-equal goal. The screen phase is tuned for recall, meaning that when the model is unsure whether a paper belongs in ethics or safety rather than reliability or general ML, the model chooses ethics or safety and assigns low confidence, so that a wrongly kept candidate can be removed in Stage 2, while a wrongly dropped one would be lost. Our tie-breaking rules used in our prompt instructions are detailed in Appendix~\ref{app:class-def-score-rules}. Of 69,290 papers with a usable abstract, \textbf{8,374 (12.1\%)} were screened in: \textbf{5,202 ethics (7.5\%)} and \textbf{3,172 safety (4.6\%)}.

The screening prompt gives the definitions of the four classes (Appendix~\ref{app:class-def-score-rules}) and asks the model to assign exactly one class by the paper's main focus, with a confidence level and a one-to-two-sentence justification. It states that a paper belongs in the ethics or safety class only if that concern is at least a co-equal goal, and that the model must choose ethics or safety, with low confidence, when unsure, because a wrongly dropped paper is lost from the study. It also lists the scope rules for recurring borderline cases (Appendix~\ref{app:scope-rules}). In the prompt, the classes are numbered 1 (ethics), 2 (reliability and XAI), 3 (general ML), and 4 (safety).
 
\subsection{Subdomain classification prompt}
\label{app:prompt-subdomain}
The Stage 2 prompt (about 4,100 words) receives the paper's title, abstract, and fixed Stage 1 class. It restates the class definitions, lists both subdomain taxonomies with a definition for each subdomain (Table~\ref{tab:taxonomy}), and includes worked boundary cases for differential privacy, fairness algorithms, unlearning, attacks on language models versus narrow models, and interpretability. Before coding race, the model must copy the shortest phrase from the abstract that shows how the paper treats race. Table~\ref{tab:codebook} lists the fields it returns.
 
\begin{table}[ht]
\centering
\footnotesize
\caption{Codebook: fields returned by the Stage 2 prompt. Race fields marked $^\ast$ are coded only for race-focused papers.}
\label{tab:codebook}
\begin{tabular}{@{}l p{0.68\linewidth}@{}}
\toprule
Field & Values \\
\midrule
    Class fit & Agree with the Stage 1 class, or name a better class (recorded, not applied) \\
    Motivation & 1 = side note; 2 = co-equal goal; 3 = primary goal \\
    Subdomain & One primary and an optional secondary subdomain from the class's taxonomy \\
    Specificity & Methodology and evaluation, each scored 0--2 \\
    Method type & Benchmark or dataset; system evaluation; training intervention; inference intervention; internal analysis; attack; formal theory; human study; conceptual or legal analysis; audit; review (multi-label) \\
    Evidence type & Automated metrics; LLM judge; human evaluation; participant study; deployed case; formal proof; argument only (multi-label) \\
    Social harm & Whether the abstract mentions harm AI causes to people or social groups (true/false) \\
    Harm object & The harm, in at most ten words, in the abstract's own terms \\
    Harm framing & Vocabulary used for the harm: safety; ethics; both; neither \\
    Race evidence & Shortest verbatim phrase showing how the paper treats race \\
    Race focus & Central; substantial; incidental (dataset attribute only); none \\
    Communities$^\ast$ & Black or African diaspora (including through African American English); other named group; unspecified minorities \\
    Harm type$^\ast$ & Representational; allocative; quality-of-service; surveillance and policing \\
    Contribution$^\ast$ & Measurement; evaluation; mitigation; governance; critique \\
    Intersectional$^\ast$ & Whether findings are intersectional (true/false) \\
    Racial-harm framing$^\ast$ & Vocabulary used for the racial harm: safety; ethics; both; neither \\
\bottomrule
\end{tabular}
\end{table}

\end{document}